\documentclass[useAMS,usenatbib,letters]{mn2e}
\usepackage{psfig}
\usepackage[dvips]{graphicx}

\newcommand{\ltorder}{\hbox{ \rlap{\raise 0.425ex\hbox{$<$}}\lower
		      0.65ex\hbox{$\sim$} }}
\newcommand{\gtorder}{\hbox{ \rlap{\raise 0.425ex\hbox{$>$}}\lower
		      0.65ex\hbox{$\sim$} }}
\newcommand{\msun}{{\ensuremath{M_\odot}}}

\newcommand{\tcc}{\mbox{${t_{\rm cc}}$}}
\newcommand{\mm}{\mbox{${\langle m \rangle}$}}
\newcommand{\mpivot}{\mbox{${m_{\rm p}}$}}
\newcommand{\rcore}{\mbox{${r_{\rm core}}$}}
\newcommand{\Wo}{\mbox{${W_0}$}}

\newcommand{\trh}{\ensuremath{t_{\rm r}}}
\newcommand{\trc}{\ensuremath{t_{\rm rc}}}
\newcommand{\rh}{\ensuremath{r_{\rm h}}}

\usepackage{aas_macros}

\def\apgt{\ {\raise-.5ex\hbox{$\buildrel>\over\sim$}}\ }
\def\aplt{\ {\raise-.5ex\hbox{$\buildrel<\over\sim$}}\ }

\title[The Arches Cluster Mass function]
      {The present day mass function in the central region of the 
       Arches cluster
      }

\author[Simon F.\, Portegies Zwart et al.]
       {Simon Portegies Zwart,$^{1,2}$\thanks{E-mail: spz@science.uva.nl (SPZ);
	egaburov@science.uva.nl (EG); huichen@science.uva.nl (H-CC);
	ato@science.uva.nl (MAG)} 
    	 Evghenii Gaburov,$^{1,2}$\footnotemark[1] 
	 Hui-Chen Chen$^{1,2,3}$\footnotemark[1] and \newauthor
	 M. Atakan G\"urkan$^{1}$\footnotemark[1] \\
$^1$    Astronomical Institute 'Anton Pannekoek'
	 University of Amsterdam,
	 the Netherlands \\
$^2$    Section Computational Science,
	 University of Amsterdam,
	 the Netherlands \\
$^3$    Graduate Institute of Astronomy, National Central
         University, No 300 Jhongda Rd. Jhongli City, Taiwan 
       }

\begin{document}

\maketitle
\date{Accepted XXX. 
      Received XXX
     }
\pagerange{\pageref{firstpage}--\pageref{lastpage}} 
\pubyear{2007}
\label{firstpage}

\begin{abstract}

We study the evolution of the mass function in young and dense star
clusters by means of direct $N$-body simulations. Our main aim is to
explain the recent observations of the relatively flat mass function
observed near the centre of the Arches star cluster. In this region,
the power law index of the mass function for stars more massive than
about 5--6\,\msun\, is larger than the Salpeter value by about unity;
whereas further out, and for the lower mass stars, the mass function
resembles the Salpeter distribution.
We show that the peculiarities in the Arches mass function can be
explained satisfactorily without primordial mass segregation. We
draw two conclusions from our simulations: 1) The Arches initial mass
function is consistent with a Salpeter slope down to $\sim 1$\,\msun,
2) The cluster is about half way towards core collapse.
The cores of other star clusters with characteristics similar to those
of the Arches are expected to show similar flattening in the mass
functions for the high mass ($\apgt 5$\,\msun) stars.
\end{abstract}


\section{Introduction}

The mass function of a star cluster changes because of
both stellar evolution and stellar dynamics.  Stellar
evolution causes the turn-off mass to decrease as the most massive
stars evolve away from the main sequence, ascend the giant branch to
ultimately shed their envelopes to turn into compact objects.  Stellar
evolution therefore has a characteristic effect on the mass function
by truncating it at the high mass end.

The dynamical evolution of a cluster has a more complicated effect on
changes in the mass function. The dominant effect here is dynamical
friction, which causes the most massive stars to sink to the cluster
centre on a time scale that is inversely proportional to the stellar
mass, i.e. the most massive stars tend to sink more quickly than 
relatively lighter stars. At the same time, stars less massive than
the average mass tend to leave the inner regions.  As a result of
this {\em mass segregation}, the local stellar population becomes a
function of the distance to the cluster centre.

Mass segregation, though mostly noticeable in the cluster's central
regions, is a global phenomenon. A star cluster that is born with the
same mass function across its radial coordinate will gradually grow a
top-heavy mass function in its centre and a top-depleted mass function
in its outskirts.  Near the half mass radius, the mass function
remains closest to the initial mass function
\citep{1997MNRAS.289..898V}.

In this letter, we concentrate on the evolution of the stellar mass
function in the inner part of young and dense star clusters, using
$N$-body simulations. Our
interest in this topic was initiated by the recent accurate
measurements published by
\cite{2005ApJ...628L.113S,2006ApJ...653L.113K} in which the mass
function in the inner $\sim 10$'' from the centre of the Arches star
cluster was studied. 
These observations, especially the latter, revealed that the mass
function of near the centre of Arches cluster is a broken power law,
with the turning point $m_{\rm p} \sim 5{\rm -}6 M_\odot$. We were
able to reproduce this feature without invoking any special mechanism.
Our simulations allow us to draw conclusions on the history of
the dynamical evolution of the Arches cluster.

\section{Dynamical evolution of the mass function}
\subsection{Parameters for the simulations}
As a cluster evolves, stars more massive than the mean mass \mm\,
tend to sink to the cluster centre whereas lighter stars move
outwards. For the most massive stars, the time scale for dynamical
friction is proportional to two-body relaxation time, $t_{\rm r}$:
\begin{equation}
t_{\rm df} \propto {\mm \over m_\star} t_{\rm r},
\end{equation}
were $m_\star$ is the mass of the massive star, which segregates inwards.
The value of the relaxation time at the cluster's half-mass radius,
$r_{\rm h}$ is given by \citep[][eq.~2.63]{Spitzer87}
\begin{equation}
  \trh = {0.138 N \over \ln \Lambda} \left({ \rh^3 \over G M
               }\right)^{1/2}.
\label{Eq:trlx}\end{equation}
Here $G$ is Newton's constant of gravity, 
$M$ and $N$ are the total mass and the number of stars in the
cluster and $\ln \Lambda$ is the Coulomb
logarithm, for which we adopt $\ln \Lambda = \ln(0.01N)$
\citep{1996MNRAS.279.1037G}. For the central relaxation time, we use
\begin{equation}
\label{eq_trc}
      t_\mathrm{rc} = \frac{ \sigma_{\rm 3D}^3 }
                           { 4.88\pi \,G^2 \ln\Lambda\,n \langle m_c \rangle^2 },
\end{equation}
where $\sigma_{\rm 3D}$, $n$ and $\langle m_c \rangle$ are the
three-dimensional velocity dispersion, number density and average
stellar mass at the cluster centre \citep[][eq.~3.37]{Spitzer87}.

We follow the dynamical evolution of our models by means of direct $N$-body
simulations, which we carry out with the {\tt starlab} software
environment \citep{2001MNRAS.321..199P}. The calculations are performed
on the GRAPE-6 special purpose computer
\citep{1997ApJ...480..432M,2001ASPC..228...87M}.

Our numerical experiments are performed with $N=12288$ and 24576
stars. For each $N$, we perform simulations starting with a full range
of density profiles for which we chose \cite{1966AJ.....71...64K}
models with the dimensionless parameter $\Wo$ ranging from 3 to 12.
The mass function in our simulations is described by a power-law,
$dN/dm = m^x$, where we adopt the Salpeter value for the index
($x=-2.35$), with masses ranging from 1\,\msun\, to 100\,\msun.  To
validate our results, we carried out additional simulations with
$N=49152$ as well as with a Salpeter mass function with 0.1\,\msun\,
as the lower limit. It will turn out that the presence of a tidal
field has little effect on the results, but reducing the lower limit
to the initial mass function to 0.1\,\msun\, has a profound effect on
the results, as we discuss below.  For clarity we mainly focus on the
models with 12288 and 24576 stars. With these parameters, the
relaxation time at the virial radius for the 12k\,models is about 360
$N$-body time units, whereas for the 24k models this is 625 $N$-body
time units (see \cite{1986LNP...267..233H}\footnote{For the definition
of an N-body unit, or the summary at {\tt
http://en.wikipedia.org/wiki/Natural\_units}.}).

The close proximity of the Arches cluster to the Galactic centre
\citep{1992AAS...181.8702C,1996ApJ...461..750C} would seemingly
require the simulations to include tidal effects.  And for
understanding the dynamics in the cluster outskirts or the evaporation
time scale the tidal field will prove crucial.  For studying the
evolution of the central region on the short time scale reported here,
however, the tidal field has negligible effect.  We support this
statement by carrying out additional simulations which include the
tidal field, and those show no discernible effect.  We therefore focus
on the results of simulations without a tidal field. This has the
attractive side effect that it allows us to scale our results with
respect to $N$. We also ignore the effects of stellar evolution.  This
approximation is rectified as on the short lifetime of the cluster
($2\pm 1$\,Myr) even the most massive stars remain on the main
sequence, though some effect of the stellar mass loss at the top end
of the mass function can be expected.  For example, a 60\,\msun\,
zero-age main sequence star with solar metalicity loses about
3\,\msun\, in its first $\sim 2.4$\,Myr \citep{2001A&A...366..538L},
which has a negligible effect on the slope of the mass function.

\subsection{Dynamical evolution towards core collapse}

\begin{figure}
\begin{center}
\psfig{figure=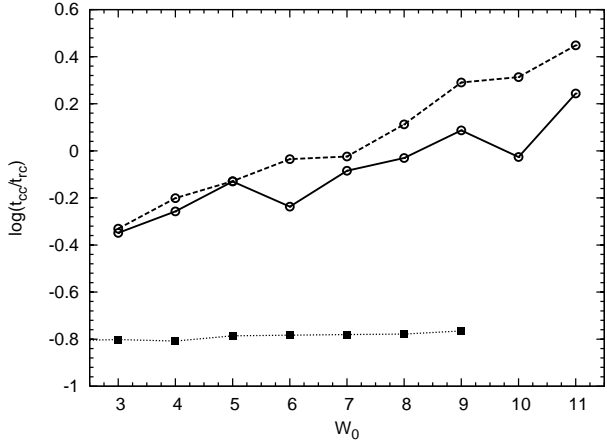,width=\columnwidth}
\end{center}
\caption{The moment of core collapse (\tcc) in units of the
         core-relaxation time (\trc) as a function of $W_0$.  The
         dashed curve gives the results of our simulations with 12k
         stars and the solid curve with 24k. Since we performed only
         one simulation per set of initial conditions no error bars
         are presented. The bottom (dotted) line denote the results of
         \protect\cite{2004ApJ...604..632G}.  }
\label{fig:tcc}
\end{figure}

In our simulations we identify the moment of core collapse as soon as
a persistent binary forms with a binding energy of at least 100\,kT
(where the energy scale kT is defined by the condition that the total
stellar kinetic energy of the system, excluding internal binary
motion, is $\frac32N$kT). For a cluster with a mass function that is
consistent with the observed mass function in young star clusters,
core collapse occurs at a more or less constant fraction of the
initial central relaxation time $t_{\rm cc} \sim 0.2\pm0.1\trc$
\citep{2002ApJ...576..899P,2004ApJ...604..632G}.  In
Fig.\,\ref{fig:tcc} we plot the moment of core collapse as a function
of the initial concentration of the cluster.  The slight dependence of
$\tcc/\trc$ on $W_0$, as well as the offset between our results with
those of \cite{2004ApJ...604..632G} is presumably mainly caused by
their broader range of stellar masses ($0.2 < m/\msun < 120$) in the
initial mass function, whereas here we adopt an initial mass function
with $1 < m/\msun < 100$. An additional effect is expected from
the difference in the number of stars. The simulations of
\cite{2004ApJ...604..632G} were carried out with $10^6$.  The
systematic difference between our results from simulations with 12k
and 24k stars shows that this also affects the results
systematically. At this moment, however, we cannot quantify this
effect.

\begin{figure}
\begin{center}
\psfig{figure=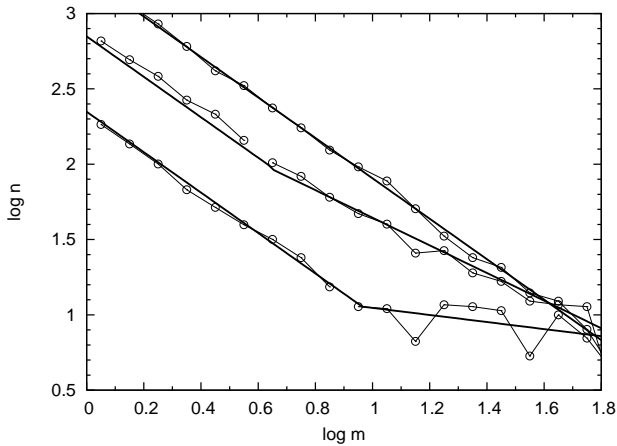,width=\columnwidth}
\end{center}
\caption{The mass function of one of the simulations ($N=24k$,
$W_0=5$) for an annulus $1<r/\rcore<2$, around the cluster
centre. From top to bottom, the curves are at times $t=0$, $t \simeq
0.61\tcc$, and $t \simeq 1.05\tcc$.
The solid lines are least squares fits to the mass function with a broken
power-law.
}
\label{fig:MFvstime}
\end{figure}

Dynamical friction causes the massive stars to segregate to the
cluster centre making the mass function flatter at the higher end, in
this region, until the formation of a hard binary.  In
Figure\,\ref{fig:MFvstime} we illustrate the evolution of the mass
function between \rcore\, and 2\rcore. We show the mass function at
birth (top curve, the Salpeter mass function), halfway to core
collapse and at core collapse (bottom curve).  We denote the point
around which the slope of the mass function changes by \mpivot, and
the power-law indices in higher and lower ends by $x_{\rm m<m_p}$ and
$x_{\rm m>m_p}$, respectively. The apparent decrease in the number of
stars in the mass function presented in Figure\,\ref{fig:MFvstime} is
the result of the cluster becoming more concentrated which causes the
adopted annulus ($\rcore < r < 2\rcore$) to become narrower.

\begin{figure}
\begin{center}
\psfig{figure=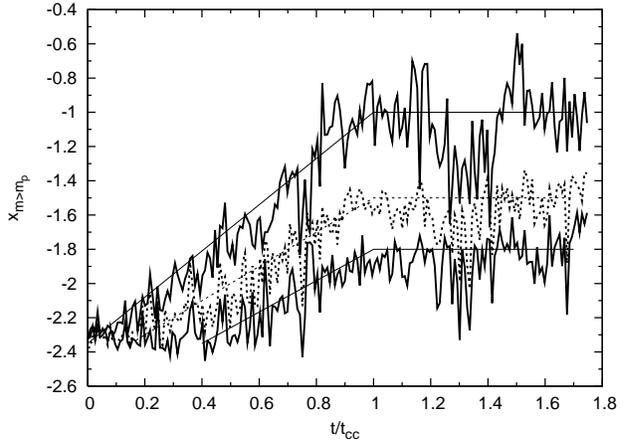,width=\columnwidth}
\end{center}
\caption{The evolution of $x_{\rm m>m_p}$ for various radial bins in
         the 24k simulation with a $W_0 = 5$ King model.
	 The time is given in units of core collapse time \tcc.
	 The radial bins are $0<r/\rcore<1$ (upper solid curve),
	 $1<r/\rcore<2$ (dotted curve), and $2<r/\rcore<3$
         (lower solid curve). To guide the eye, we plotted
         straight lines through the simulation data.  }
\label{fig:xhm}
\end{figure}

The effect of flattening of the mass function is less pronounced
further away from the cluster centre. This is illustrated in
Figure\,\ref{fig:xhm}, where we present the evolution of $x_{\rm
m>m_p}$ for the 24k simulations with $W_0 = 5$ and for $r=0$ to
\rcore\, (top curve), for $r=\rcore$ to 2\rcore\, and for $r=2\rcore$
to 3\rcore\, (bottom curve).  The values of $m_{\rm p}$, $x_{\rm
m<m_p}$ and $x_{\rm m>m_p}$ are obtained by a three-point least
squares fit to the mass function in a predetermined annulus of the
simulated date. Note that we relaxed the fitting procedure in the
sense that the mass function is not required to be continuous.  The
point of stalling of the evolution of the mass function can be
identified by the moment of core collapse, regardless of the initial
concentration or the number of stars in the simulation.  Therefore, we
normalize the time axis in Figure\,\ref{fig:xhm} to that instant.

\begin{figure}
\begin{center}
\psfig{figure=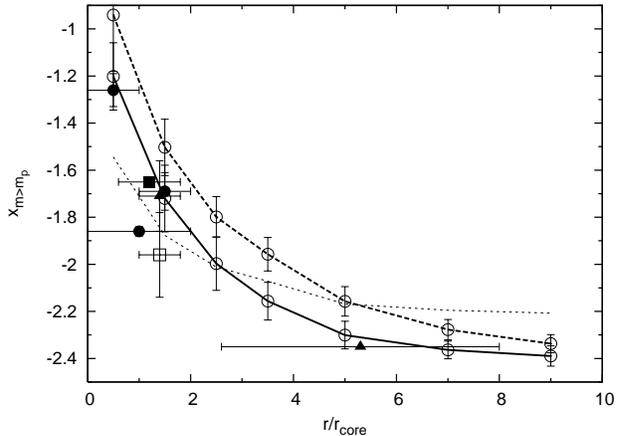,width=\columnwidth}
\end{center}
\caption{The value of $x_{\rm m>m_p}$ as a function of distance form
         the cluster centre after core collapse.  The solid curve
         gives the average value of $x_{\rm m>m_p}$ over the various
         simulations with 24k stars for $W_0=3$, 4 up to $W_0=12$, the
         dashed line gives the data for the simulations with $N=12$k.
         The squares, bullets and triangles with error bars give the
         observed values taken from
         \protect\cite{1999ApJ...525..750F},
         \protect\cite{2005ApJ...628L.113S} and
         \protect\cite{2006ApJ...653L.113K}, respectively (see
         Tab.\,\ref{Tab:Arches}).
The lower thin dashed line gives the value of $x_{\rm m>m_p}$ for
simulations with 24k particles with $W_0=9$ and a lower limit to the
initial mass function of 0.1\,\msun.  }
\label{fig:xmgtmp}
\end{figure}

\begin{figure}
\begin{center}
\psfig{figure=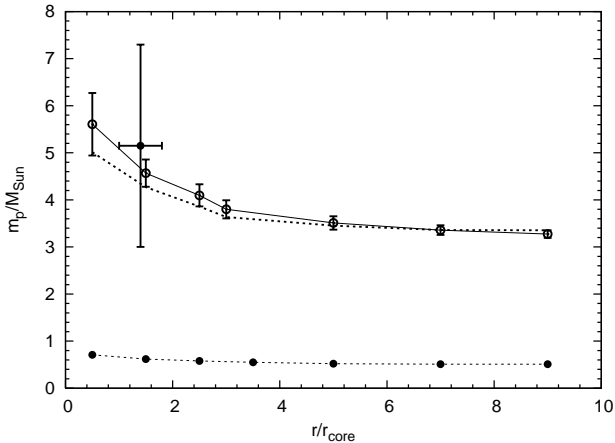,width=\columnwidth}
\end{center}
\caption{The value of \mpivot\, after the point of core collapse, as a
         function of distance to the cluster centre for various of
         simulations.  The circles connected with a thin solid line
         give the results for the simulations with 24k stars and with
         $W_0=5$ averaged between the moment of core collapse and
         twice the core collapse time. The error bars indicate the
         variation of the value of \mpivot\, over this time period.
         More concentrated initial models tend to have a slightly
         lower value of \mpivot, which we illustrate by plotting the
         $W_0=9$ simulation as the thick lower dotted line.
         The upper dotted line gives the results of the simulation
         with $N=12k$ and for $W_0=5$.
         The thin dashed line at the bottom gives the value of
         \mpivot\, for simulations with 24k particles with $W_0=9$ and
         a lower limit to the initial mass function of 0.1\,\msun.  }
\label{fig:mp_vs_r}
\end{figure}

\subsection{Post-collapse mass function}
After the formation of a hard binary, the mass function achieves a
quasi steady state. The slope of the high-mass end of the mass
function varies throughout the cluster.  In in
Figure\,\ref{fig:xmgtmp} we show how the mass function for stars with
$m > m_p$ after the moment of core collapse is a function of the
distance to the cluster centre, being flatter closer in and resembling
the initial mass function further out. Overplotted are the observed
values of the mass function exponent (see Tab.\,\ref{Tab:Arches}, see
\S\,\ref{Sect:Arches}). The results of our simulations with a tidal
field are statistically identical to those with a tidal field. The
simulations with a minimum mass to the initial mass function of
$m_{\rm min} = 0.1\,\msun$\, is plotted as the thin dashed curve in
Fig.\,\ref{fig:xmgtmp}. For clarity we did not plot error bars for
this figure, but the results with a lower limit of 0.1\,\msun\, are
inconsistent with the observed values.

In Figure\,\ref{fig:mp_vs_r}, we show the value of \mpivot\, as a
function of distance from the cluster centre for various simulations,
past the moment of core collapse. It turns out that more concentrated
initial models tend to result in a slightly smaller value of \mpivot\,
whereas simulations with a smaller number of stars give rise to a
higher value of \mpivot.  The behaviour of $x_{\rm m>m_p}$ is rather
insensitive to the initial concentration of the cluster.

It may be noted that the results of simulations with 0.1\,\msun as the
lower limit of the mass function are not consistent with the observed
values of $x_{\rm m>m_p}$ and $m_p$.

\section{Mass function of the Arches cluster}\label{Sect:Arches}
\label{Sect:comparison}
   
At a projected distance of about 25\,pc from Sgr A*, the Arches
cluster ($\alpha = 17^h45^m50^s$, $\delta = -27^\circ49'28''$ in
J2000), discovered by \cite{1992AAS...181.8702C,1996ApJ...461..750C},
is peculiar. The total cluster mass is about $2\cdot 10^4$\,\msun\,.
The core radius of the cluster (defined as the radial distance from
the cluster centre where the luminosity profile drops by a factor two)
is \rcore = 5''.0 \citep{2005ApJ...628L.113S}, and corresponds to
about 0.2\,pc if we assume that the distance to the Galactic centre is
8\,kpc.  The age of the cluster is $2 \pm 1$\,Myr
\citep{1999ApJ...525..750F}.

Recently, \cite{2006ApJ...653L.113K} observed the Arches cluster using
Keck/NIRC2 laser guide star adaptive optics. Their observations covered
the inner parts of the cluster and some control fields at a distance of about
2.4\,pc (60'') from the cluster centre. They subsequently constructed
the luminosity and mass functions down to about 1.3\,\msun, in an
annulus of 5'' (about 1.0\,\rcore) to 9'' (about 1.8\,\rcore) from
the cluster centre. In Table\,\ref{Tab:Arches} we give the various
measurements of the mass function and the distance from the cluster
centre in terms of the observed core radius.
\begin{table}
\caption[]{
   Parameters for the observed mass function of the Arches cluster.
   The first two columns give the range over which the mass function
   is measured, in units of the cluster's core radius ($\rcore \simeq
   0.20$\,pc).  The third and fourth columns give the range in masses
   for which the exponent of the mass function (last column) is
   fitted.  Column 5 gives the reference for the mass function
   exponents 1: \cite{2005ApJ...628L.113S}, 2:
   \cite{2006ApJ...653L.113K}, 3: \cite{1999ApJ...525..750F}, and the
   last column the measured value of $x$ between $m_{\rm min}$ and
   $m_{\rm max}$.
}
\label{Tab:Arches}
\begin{tabular}{lrrrrc}
$r_{\rm min}$ &$r_{\rm max}$&$m_{\rm min}$ &$m_{\rm max}$&ref& $x$ \\
(\rcore)      & (\rcore)     & (\msun)   &(\msun)      &   &  \\
\hline
0     &  1    & 12           & 60    &1& $-1.26\pm0.07$ \\
1     &  2    & 6            & 16    &1& $-1.69\pm0.08$ \\
1     &  2    & 16           & 60    &1& $-2.21\pm0.09$ \\
0     &  2    & 6            & 60    &1& $-1.86\pm0.02$ \\
1     &  1.8  & 6.3          & 50    &2& $-1.71\pm0.15$ \\
1     &  1.8  & 1.3          & 50    &2& $-1.91\pm0.08$ \\
0.6   &  1.8  & 6.3          & 125   &3& $\sim -1.65$ \\
4     &  8    & 2.8          & 32    &3& ${\cal O}(-2.35)$ \\
\end{tabular}
\end{table}

\begin{figure}
\begin{center}
\psfig{figure=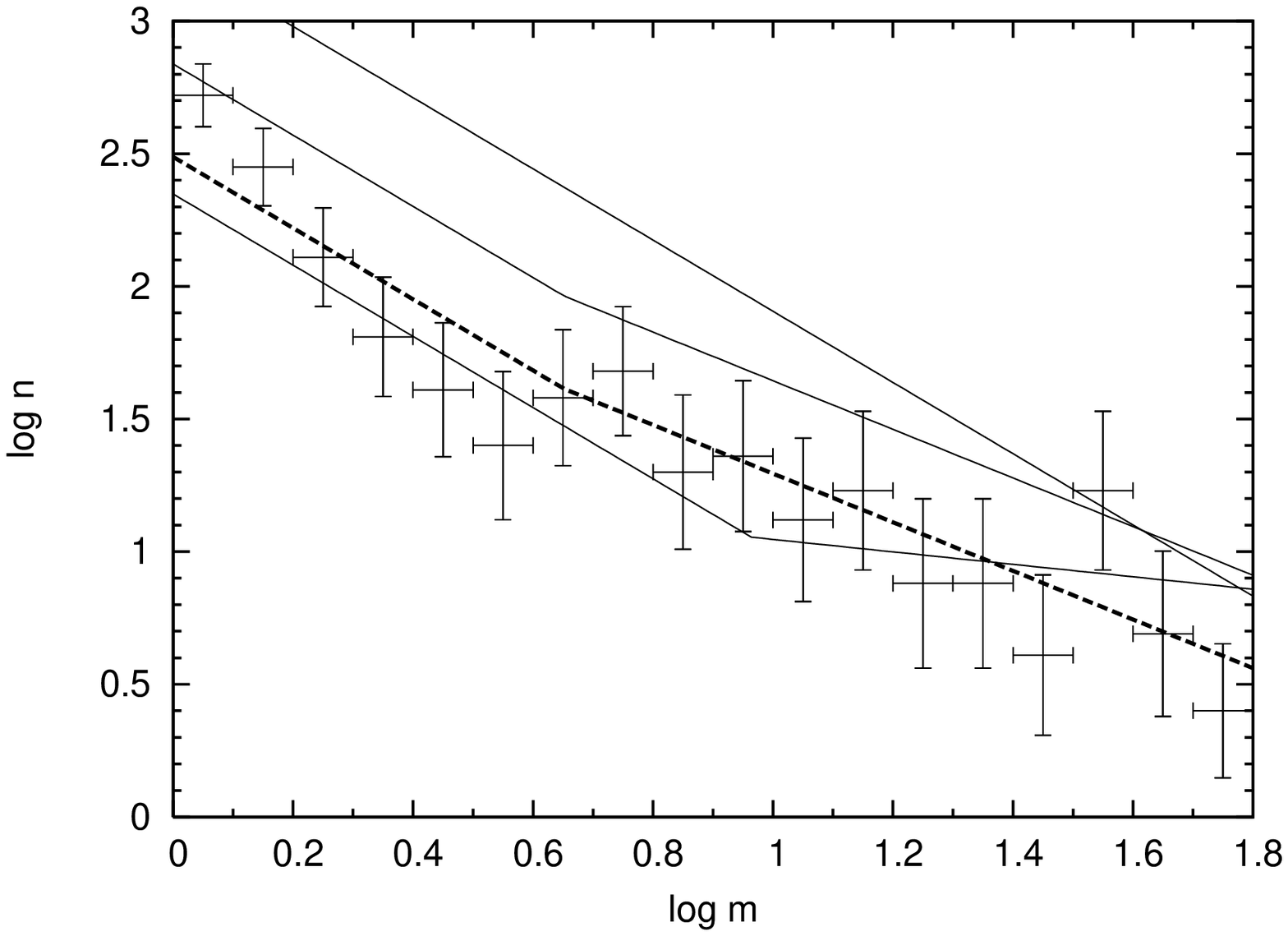,width=\columnwidth}
\end{center}
\caption{ The present day mass function of the Arches star cluster
between 1\,\rcore\, and 1.8\,\rcore.  The error bars are taken
directly from Fig.\,5 of \protect\cite{2006ApJ...653L.113K}.  The three thin
lines are from Fig.\,\ref{fig:MFvstime} at zero age
(top line), at $t \simeq 0.61\tcc$ (middle line) and at $t \simeq
1.05\tcc$ (bottom line). The thick dashed line is identical to the thin
$t \simeq 0.61\tcc$ but then renormalized with -0.35 dex. This line
produces a satisfactory fit to the observed mass function.
}
\label{fig:comparison}
\end{figure}

The data show that the slope in the mass function for stars more
massive than $\sim 5$\,\msun\, flattens towards the cluster centre
(see fig.\,\ref{fig:xmgtmp}). For lower mass stars, as well as for
further out than $\sim 4$\,\rcore\, it is closer to to the Salpeter
mass function. The observed mass function in the Arches is not a
simple power-law \citep{2005ApJ...628L.113S,2006ApJ...653L.113K}.  We
argue that in the inner parts of the cluster, $r\aplt 4\rcore$, the
mass function is best described by two power laws with the break
around $m_p = 5$--6\,\msun. The mass function below this break ($m_p$)
resembles the initial mass function ($x_{\rm m<m_p} \equiv x_{\rm
  IMF}$), and above $m_p$ it becomes flatter ($x_{\rm m>m_p} > x_{\rm
  IMF}$). Further out than $r \simeq 4\,\rcore$\, the break disappears
and the mass function becomes gradually better represented with the
initial mass function.

\section{Discussion and conclusions}

We performed detailed simulations of the evolution of young and dense
star clusters using direct $N$-body simulations, in order to constrain
the observed mass function within about one parsec from the centre of
the Arches cluster.  The initial conditions of our simulations range
over the full spectrum of King model density profiles.  

The mass function in the central region of the Arches cluster is
peculiar as it appears to be split in two power-laws, one for the
stars less massive than 5--6\,\msun\, and a much shallower slope for
the more massive stars.  The two power-laws fit the observed data
between 5'' and 9'' is marginally better ($\chi^2 \simeq 0.5$) than a
single power-law ($\chi^2 \simeq 0.93$).

The simulations we perform to mimic the Arches cluster are able to
reproduce this observed broken power-law mass function at the observed
projected distance from the cluster centre ($r=0$--4\,\rcore).  The
best comparison between observations and simulations is obtained if
the cluster is about half way core collapse ($t = 0.4$--0.6\,\tcc).

Our simulations, however, are performed without stellar evolution and
without including the effects of an external tidal potential. As a
result, they are scale-free, and no specific choices for the scalings
to mass, size and therefore to time are obliged.  However, the
scale-free aspect of our simulations hinders the direct comparison to
some extend as the size scale (in parsec) and time scale (in Myr) are
important for an unbiased comparison with the observed Arches cluster.

In the comparison with the observations we adopt the same definition
of the core radius by projecting the cluster and assigning
luminosities of all stars in our simulations using zero-age
main-sequence luminosities\footnote{For comparison, we adopt the
  observers' definition of core radius: the point where the surface
  brightness drops to half its central value. This is similar to but
  different from definitions used in theoretical works of
  \cite[eq. 1-34]{Spitzer87} and \cite[eq. 15-4]{2003gnbs.book.....A}.
  In starlab we adopt the method as discussed by
  \cite{2006MNRAS.368..677H}.  The latter definition of the core
  radius systematically is about twice the observers' definition.}.
We ignore here the fact that very massive stars may become brighter in
the $2\pm1$\,Myr lifetime of the cluster, but this only affects the
most massive stars, whereas the measurements are dominated by stars in
the mid-range of masses.

The Arches cluster does not show any evidence for primordial mass
segregation as our simulations (which were initialized without
primordial mass segregation) are able to satisfactorily reproduce the
observed mass function over the entire range of observed masses and
distances from the cluster centre. Note also that the presence of
primordial gas which failed to form stars does not seem to have
affected the early cluster evolution, as the observed cluster
structure at an age of $2\pm1$\,Myr is satisfactory explained with the
simulations, which do not include gas dynamics.  The initial mass
function of the Arches cluster is then consistent with a Salpeter
slope between 1\,\msun\, and 100\,\msun\, without the need for a
radial dependence. There seems to be no need for a large population of
stars less massive than $\sim 1$\,\msun.

In fig.\,\ref{fig:xmgtmp} we show the evolution of $x_{\rm m>m_p}$
for the annuli and distances from the cluster centre reported from our
compilation from the literature in Tab.\,\ref{Tab:Arches}.  The best
match between the simulations and the observations is acquired for
simulations between $t = 0.4\tcc$ and 0.6\,\tcc, i.e: we predict that
the cluster is about half way towards core collapse.  We therefore
conclude that the Arches cluster has not yet experienced core collapse
but is currently in a pre collapse stage.

In Tab.\,\ref{Tab:Arches} we have quantified the slope to the low mass
end of the mass function in 5'' to 9'' annulus of the Arches cluster
as consistent with Salpeter, whereas the naive measurement in Fig.\,5
of \cite{2006ApJ...653L.113K} would results in $x_{\rm m<m_p} = -3.67
\pm 0.14$ (with $m_p = 5$\,\msun), which is unusually steep. If this
slope would represent the intrinsic Arches initial mass function and
we adopt a minimum mass of 1\,\msun\, the observed $\sim 2600$ stars
more massive than $\sim 5$\,\msun\, in the Arches cluster would result
in a total number of more than $2\times 10^5$ stars, which is
unrealistically high.  From an observational point of view there are
good arguments that the low mass end of the mass function is
over-estimated, as it is plagued by selection effects. One of these
effects is the artificial correction of missing stars in a crowded
field and the selection of the three control fields to compensate for
the background population.  In the Keck observations these control
fields are within about 2.4\,pc from the cluster centre, which
corresponds to $\sim 12$\,\rcore. For a King model with $\Wo \apgt
5.2$ the control fields would then be located near the cluster tidal
radius. And since the cluster is about half way towards core collapse
it is conceivable that the density profile is described with a King
model with $W_0 \apgt 7$, in which case the control field are part of
the cluster halo.

Due to mass segregation the cluster outskirts will be depleted of high
mass stars and low mass stars will be overrepresented (the opposite
effect as we discussed for the core population). Correcting the mass
function in the cluster core with a population taken from near the
cluster halo will therefore result in an enormous over correction
towards the low mass stars, and consequentially result in a steepening
of the 'corrected' mass function.

One of the control fields (field B of Kim et al,
2006)\nocite{2006ApJ...653L.113K} is taken near the location where one
expects the tidal tail of the cluster in the potential of the Galaxy
to pass though. The tidal tail is, since it consists of the halo
population, also likely to be dominated by low mass stars.

Each of the effects discussed tend to steepen the lower-mass end of the
mass function, though it is not trivial to quantize the effect without
a much more detailed study.  We however, argue that the initial mass
function of the Arches cluster was probably consistent with Salpeter
over the observed mass range.  The observed break in the mass function
around 5--6\,\msun\, and the consequential flattening of the mass
function for higher masses is then the result of the dynamical
evolution of the cluster.

The initial model which is most comparable to the observed Arches
cluster has a Salpeter initial mass function between 1\,\msun\, and
100\,\msun\, and with a reasonably concentrated initial density
profile ($W_0 \apgt 4$ and $W_0 \aplt 8$).

The break in the mass function in the inner parts of the cluster (for
$r \aplt 4\rcore$) appear at $\mpivot \simeq 2\mm$, which for our
simulations is at about 5\,\msun. The break in the observed mass
function in the Arches cluster appears around the same mass of
$\mpivot \simeq 5$--6\,\msun.  We performed additional simulations
with $\Wo=9$ using a Salpeter mass function down to 0.1\,\msun, and in
this case the break in the core mass function also developed around
$\mpivot \simeq 2\mm$, which for the adopted mass function is about
1.0\,\msun (Figure\,\ref{fig:mp_vs_r}). Based on these findings we
argue that the initial mass function in the Arches cluster has a lower
limit of about 1.0\,\msun, as in our simulations that reproduce the
observations best.

We predict that other clusters with similar parameters as the Arches
cluster, like Westerlund 1 \citep{1998A&AS..127..423P}, NGC 3603
\citep{2004AJ....128.2854M}, R\,136 \citep{1982MNRAS.200P...1M} and
Quintuplet \citep{1992AAS...181.8702C} will show similar
characteristics as Arches. The mass functions in their cores will also
be rather flat for stars more massive than 5--6\,\msun. And the mass
functions further away from the cluster centre will gradually be more
like the initial mass function.

\section*{Acknowlegments}

We are grateful to Peter Anders, Mark Gieles, Alessia Gualandris,
Douglas Heggie and Henny Lamers for many discussions.  This work
was supported by NWO (grants \#635.000.303 and \#643.200.503),
NOVA, the LKBF, the ISSI in Bern, Switzerland and the Taiwanese
government (grants NSC095-2917-I-008-006 and NSC95-2112-M-008-006).
MAG is supported by a Marie Curie Intra-European Fellowship under
the sixth framework programme.  The calculations for this work were
done on the MoDeStA computer in Amsterdam, which is hosted by the
SARA supercomputer centre.

\label{lastpage}
\end{document}